\documentclass{article}

\usepackage{arxiv}

\usepackage[utf8]{inputenc} 
\usepackage[T1]{fontenc}    
\usepackage{hyperref}       
\usepackage{url}            
\usepackage{booktabs}       
\usepackage{amsfonts}       
\usepackage{nicefrac}       
\usepackage{microtype}      
\usepackage{lipsum}		
\usepackage{graphicx}
\usepackage{natbib}
\usepackage{doi}

\usepackage{amsmath}
\usepackage{enumitem}
\usepackage{tcolorbox}
\usepackage{lipsum} 
\usepackage{float}
\usepackage{times}
\usepackage{latexsym}
\usepackage{hyperref}

\usepackage{multirow}
\usepackage{booktabs}
\usepackage{graphicx}
\definecolor{lightred}{rgb}{1.0, 0.5, 0.5}
\usepackage{xcolor}

\usepackage[ruled,vlined]{algorithm2e}
\usepackage{amsmath}
\usepackage{booktabs}
\usepackage{siunitx}
\usepackage{tikz}
\usepackage{pgfplots}
\usepgfplotslibrary{polar}
\pgfplotsset{compat=1.18}

\title{A Reproducibility Analysis of PO4ISR: Diagnosing and Mitigating Semantic Drift in LLM-Based Session Recommendation}

\author{ {Aditya Tiwari}\thanks{Both authors contribute equally.} \\
	MATRA Lab \\ Department of Computer Science and Engineering\\
	Indian Institute of Technology Bhilai\\
	Chhattisgarh, India 491002 \\
	\texttt{aadi.tiwari0208@gmail.com} \\
	\And
	{Konduri Naga Lakshmi Rekha}* \\
	MATRA Lab \\ Department of Computer Science and Engineering\\
	Indian Institute of Technology Bhilai\\
	Chhattisgarh, India 491002 \\
	\texttt{kondurinaga@iitbhilai.ac.in} \\
    \And
    \href{https://orcid.org/0000-0002-0096-2440}{\includegraphics[scale=0.06]{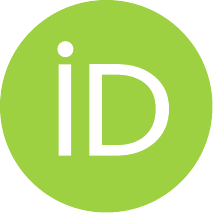}\hspace{1mm}{Rajesh Kumar Mundotiya}} \\
    MATRA Lab \\ Department of Computer Science and Engineering\\
	Indian Institute of Technology Bhilai\\
	Chhattisgarh, India 491002 \\
	\texttt{rajeshkm.mundotiya@gmail.com; rmundotiya@iitbhilai.ac.in} \\
}

\renewcommand{\shorttitle}{A Reproducibility Analysis of PO4ISR: Diagnosing and Mitigating Semantic Drift in LLM-Based Session Recommendation}

\begin{document}
\maketitle






\begin{abstract}
Reasoning-based Large Language Models (LLMs) like PO4ISR have set new benchmarks in session-based recommendation. However, the reproducibility of their reasoning capabilities across diverse semantic domains remains unexplored. In this work, we conduct a rigorous reproducibility study of PO4ISR to assess its generalization limits. Our analysis reveals a critical failure mode: standard reasoning prompts suffer from severe contextual drift in long sessions, leading to performance degradation on semantically complex datasets like Games and Bundle.

To quantify and resolve this stability gap, we introduce PO4ISR++, a robustness-enhanced implementation that integrates reflexive prompting and consistent rank detection. Unlike the original static prompting strategy, our approach dynamically adapts to cross-domain cues. We benchmark both the original implementation and our robust variant on ML-1M, Games, and Bundle. Our results confirm that while the original model struggles in new domains, our reproducible extension restores performance, yielding a stabilized gain of up to 54\% on Games and 96\% on Bundle. We release open-source artifacts, including the reproduced baseline and our enhanced framework, to facilitate reliable future research in LLM-based recommendation.
\end{abstract}



\maketitle

\section{Introduction}

Session-based recommendation (SR) has established itself as a cornerstone of modern information retrieval, adhering to the principle of privacy by predicting user preferences based solely on short-term interaction sequences~\cite{liu2018stamp,li2017neural,yuan2019simple}. Unlike conventional collaborative filtering, which relies on persistent user profiles, SR systems must infer evolving user intent from sparse, transient behavioral cues-a challenge critical for anonymous environments in e-commerce and media streaming.

While early approaches leveraged Recurrent Neural Networks (RNNs)~\cite{hidasi2015session,tan2016improved} and Graph Neural Networks (GNNs)~\cite{qiu2019rethinking,wu2019session} to model sequential dependencies, recent advancements have shifted toward the reasoning capabilities of Large Language Models (LLMs)~\cite{hou2024large,zhang2025recommendation}. Notable frameworks such as PO4ISR~\cite{sun2024large} demonstrate that LLMs can extract deep semantic insights via prompt engineering, offering a promising avenue for interpretable recommendation.

However, the transition from latent vectors to prompt-based reasoning introduces new barriers to system reliability and reproducibility. In reproducing state-of-the-art LLM-based approaches, we identify two critical failure modes that hinder practical deployment:
\begin{enumerate}
    \item \textbf{Stochastic Output Instability:} Current methods often rely on text-generation to predict item indices. We observe that this approach degrades significantly when item metadata contains numerical ambiguity (e.g., \textit{"Goya Foods Mango Nectar, 5-Ounce [Pack of \textbf{48}]"} or \textit{"Xbox \textbf{360}"}). The LLM frequently misinterprets these quantities as ranking indices, leading to hallucinated outputs and parsing failures that sever the link between reasoning and recommendation.
    \item \textbf{Domain Overfitting via Static Prompting:} Existing strategies typically employ a ``Top-1'' prompt optimization approach~\cite{sun2024large}, where a prompt tuned for a specific dataset is applied universally. Our analysis reveals that this results in domain-specific overfitting, failing to generalize to datasets with distinct semantic structures (e.g., moving from Movies to Bundles).
\end{enumerate}

These limitations highlight a gap between the theoretical potential of reasoning-based SR and its reproducible robustness across diverse real-world domains.

\subsection{Contributions}
To bridge this gap, we present a reproducibility study and robustness enhancement of the reasoning-based SR paradigm. We introduce \textbf{PO4ISR$^{++}$}, a reflexive deterministic framework designed to ensure consistent performance across domains. Our specific contributions are as follows:

\begin{itemize}
    \item \textbf{Diagnostic Reproducibility Analysis:} We quantify the stability gap in existing LLM-SR models, systematically categorizing failure cases caused by numerical ambiguity and prompt stochasticity.
    \item \textbf{Deterministic Rank-Mapping:} We propose an index-specific output formatting strategy that decouples semantic reasoning from syntax generation. This eliminates parsing ambiguity, ensuring valid, reproducible rankings even for items with complex numerical nomenclature.
    \item \textbf{Reflexive Cross-Domain Fusion:} Leveraging the meta-reasoning capabilities of Gemini-2.0, we introduce a reflexive prompting strategy. Unlike static methods, this approach fuses insights from multi-domain training (ML-1M, Games, Bundle) to discover generalizable prompt patterns that transfer effectively across disparate semantic contexts.
    \item \textbf{Robustness Benchmarking:} Extensive experiments demonstrate that PO4ISR$^{++}$ not only resolves the parsing failures of its predecessor but also achieves statistically significant performance gains, establishing a new, robust baseline for the reproducible evaluation of LLM-based recommenders.
\end{itemize}

\section{Related Work}

\subsection{Evolution of Session-Based Recommendation}
Session-Based Recommendation (SR) has progressed through three distinct modeling paradigms: sequential, graph-based, and attention-driven architectures. Early sequential approaches pioneered by Hidasi et al.~\cite{hidasi2015session} utilized Gated Recurrent Units (GRUs) to model interaction sequences. Subsequent variants integrated attention mechanisms (NARM~\cite{li2017neural}, STAMP~\cite{liu2018stamp}) to capture transient user interests. However, these recurrent architectures often struggled with long-range dependencies and the rigid ordering of interactions. To address these structural limitations, Graph Neural Networks (GNNs) emerged as a powerful alternative. Wu et al.~\cite{wu2019session} introduced SR-GNN to model complex, non-consecutive item transitions, inspiring extensions that incorporated target-aware propagation~\cite{yu2020tagnn} and global context~\cite{wang2020global}. Concurrently, the success of Transformers in NLP led to the adoption of self-attention mechanisms in recommendation (SASRec~\cite{kang2018self}, BERT4Rec~\cite{sun2019bert4rec}), which enabled parallelized processing and superior modeling of global dependencies.

Despite their predictive success, these latent-based models face a common limitation: they operate as black boxes, mapping behaviors to opaque vector spaces without explicit reasoning, often failing to capture the multiplicity of conflicting user intents within a single session~\cite{li2019multi,tang2023towards}.

\subsection{Intent-Aware \& Multi-Interest Modeling}
To enhance interpretability and granularity, Intent-Aware Session Recommendation (ISR) focuses on explicitly modeling the latent goals driving user behavior. Early ISR methods assumed a monolithic intent per session~\cite{chen2020handling,pan2020star}, which proved insufficient for dynamic browsing behaviors. More recent approaches, such as MCPRN~\cite{liu2023modeling} and HIDE~\cite{li2022enhancing}, employ multi-channel networks to disentangle concurrent user goals. Contrastive learning frameworks~\cite{chen2022intent,li2023multi} further refine this by maximizing the distinctiveness between different intent representations. While these methods improve the granularity of representations, they remain bound to latent embeddings, lacking the semantic transparency required for explainable recommendation.

\subsection{LLMs and Reproducibility Challenges in SR}
The integration of Large Language Models (LLMs) marks a paradigm shift from ``matching'' to ``reasoning.'' Unlike collaborative filtering, LLM-based approaches leverage vast pre-trained knowledge to perform zero-shot or few-shot recommendation (TALLRec~\cite{bao2023tallrec}, GPT4Rec~\cite{li2023gpt4rec}).

However, this shift introduces significant reproducibility and reliability challenges. Empirical studies~\cite{kang2023llms,liu2023chatgpt} indicate that while LLMs excel at general reasoning, they suffer from stochastic instability and ranking inconsistency when applied to specific recommendation tasks. Geng et al.~\cite{geng2022recommendation} attempted to mitigate this via unified pretrain-prompt-predict frameworks, yet domain adaptation remains a bottleneck.

Most relevant to our work is PO4ISR~\cite{sun2024large}, which established a state-of-the-art baseline by applying prompt optimization to ISR tasks. While PO4ISR demonstrated the efficacy of reasoning-based recommendations, its original evaluation relied on static, domain-specific prompts and unstructured generation. As we demonstrate in this study, these design choices lead to fragile generalization--specifically, vulnerability to numerical ambiguity and domain drift. Our work directly addresses these gaps, moving beyond simple application to rigorously benchmark and enhance the reproducibility of reasoning-based ISR.

\section{PO4ISR$^{++}$: Enhancing Reproducibility via Determinism and Fusion}

\subsection{Problem Formulation}
We frame SR as a prompt optimization task with a ranking objective. Let $\mathcal{S}$ denote a set of user sessions, where each $s \in \mathcal{S}$ is a sequence of interactions. Given a candidate pool $\mathcal{C}$ and a ground-truth next item $G(s)$, the objective is to learn an optimal prompt $P^*$ that maximizes the rank of $G(s)$ in the LLM's output distribution. Formally, we seek a mapping $\mathcal{F}: (\mathcal{S}, \mathcal{C}, G) \mapsto P^*$ such that $P^*$ generalizes across diverse domains $\mathcal{D}$, robust to variations in item nomenclature and domain-specific semantics. 

\subsection{Baseline: The PO4ISR Framework}
The original PO4ISR~\cite{sun2024large} framework acts as our primary baseline. It treats recommendation as a three-stage pipeline: (1) Error Identification, where the model ranks items using an initial prompt; (2) Reason Generation, where the LLM critiques its own ranking errors; and (3) Prompt Optimization, where these critiques are used to refine the prompt via an iterative Upper Confidence Bound strategy highlighted in the original PO4ISR~\cite{sun2024large}. While effective in theory, our reproducibility analysis reveals that this pipeline suffers from two critical stability flaws. First, it exhibits parsing fragility, wherein the reasoning loop fails when the LLM’s textual output cannot be deterministically mapped to predefined item indices. Second, it demonstrates domain myopia: the ``Top-1'' prompt selection strategy tends to overfit the dominant patterns of a single training domain, thereby limiting its ability to generalize and transfer effectively to new or heterogeneous contexts.

\begin{figure*}[t]
  \centering
  \includegraphics[width=\linewidth]{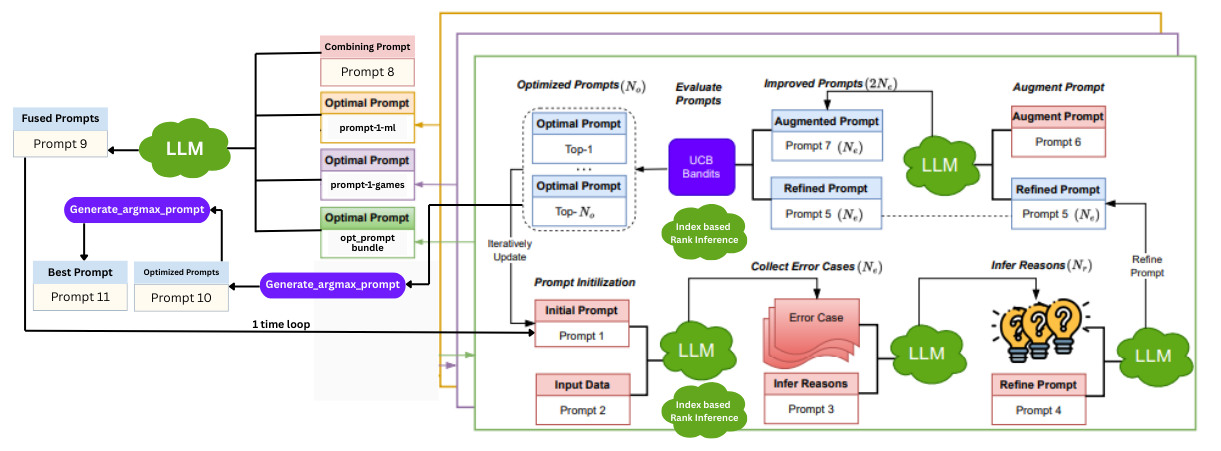}
  \caption{The overall architecture of our PO4ISR$^{++}$ framework, incorporating index-based output formatting and cross-domain prompt fusion.}
  \label{fig:architecture}
\end{figure*}

\subsection{PO4ISR$^{++}$: Robustness Enhancements}
To resolve these reproducibility barriers, we introduce PO4ISR$^{++}$. As illustrated in Figure~\ref{fig:architecture}, our framework retains the reasoning core of PO4ISR but replaces its fragile components with a Deterministic Output Interface and a Reflexive Cross-Domain Fusion module. All the prompts are listed in Appendix~\ref{sec:prompt}. 

\subsubsection{Deterministic Index-Based Output}
In the original framework, the LLM is instructed to output item names (e.g., \textit{"1. Item A"}). This introduces severe ambiguity when item names contain numerals (e.g., \textit{"Sony WH-1000XM4"}), causing the parser to mistake metadata for ranking positions. To ensure strictly reproducible evaluations, we reformulate the output space. We enforce a Zero-Based Indexing constraint, requiring the LLM to return a flat list of indices $\hat{\pi} = [\text{idx}_1, \text{idx}_2, \ldots, \text{idx}_{|\mathcal{C}|}]$, where each $\text{idx}_i \in \{0, 1, \ldots, |\mathcal{C}|-1\}$ corresponds to the item's position in the candidate list. This decoupling of semantic reasoning (internal) from syntax generation (output) eliminates numerical hallucination errors, as shown in our experimental analysis.

\subsubsection{Reflexive Reflexive Cross-Domain Fusion}
To address domain overfitting, we abandon the ``single-best prompt'' strategy. We propose a Reflexive Cross-Domain Fusion mechanism that synthesizes invariant reasoning patterns from multiple domains (ML-1M, Games, Bundle). This process operates in three stages:

\begin{enumerate}
    \item \textbf{Prompt Distillation:}
    We first identify the highest-performing domain-specific prompts (e.g., $P_{\text{ML}}^*$, $P_{\text{Games}}^*$) using the standard PO4ISR optimization loop. This isolates the distinct reasoning styles required for different item types (e.g., visual items in Games vs. bundle logic in E-commerce).

    \item \textbf{Reflexive Synthesis:}
    We concatenate these domain-specific experts and feed them into a meta-reasoner (Gemini-2.0) with the instruction to \textit{``synthesize a unified prompt that retains the visual reasoning of Games and the bundle logic of E-commerce.''} This generates a set of hybrid candidate prompts $\{P^{(1)}_{\text{fuse}}, \ldots, P^{(10)}_{\text{fuse}}\}$.

    \item \textbf{Two-Stage Verification:}
    To prevent overfitting to the fusion set, we employ a split-validation strategy (Algorithm~\ref{alg:prmp_fusion}). We fine-tune the hybrid prompts on a held-out optimization set ($D_{\text{opt}}$) comprising 15 sessions per domain, and select the final model strictly based on a separate validation set ($D_{\text{val}}$).
\end{enumerate}

The resulting prompt $P^*$ is not merely ``trained'' on multiple domains but is structurally designed to trigger different reasoning pathways depending on the input context.

\begin{algorithm}[h]
\SetAlgoLined
\DontPrintSemicolon
\caption{Reflexive Cross-Domain Fusion}
\label{alg:prmp_fusion}
\SetKwInOut{Input}{Input}
\SetKwInOut{Output}{Output}
\Input{Domain-Experts $\mathcal{P}_{experts} = \{P_{\text{ML}}, P_{\text{Games}}, \ldots\}$, Meta-Prompt $P_{\text{meta}}$}
\Output{Robust Cross-Domain Prompt $P^*$}

\Begin{
  \tcp{Step 1: Meta-Synthesis via LLM}
  $P_{\text{hybrid}} \gets \text{LLM}(\text{Concat}(\mathcal{P}_{experts}) + P_{\text{meta}})$ \\
  
  \tcp{Step 2: Split-Set Sampling}
  $D_{\text{opt}} \gets \emptyset, \ D_{\text{val}} \gets \emptyset$ \\
  \ForEach{$d \in \mathcal{D}$}{
    $D_{\text{opt}} \gets D_{\text{opt}} \cup \textsc{Sample}(d, 15)$ \\
    $D_{\text{val}} \gets D_{\text{val}} \cup \textsc{Sample}(d, 15)$ \\
  }
  
  \tcp{Step 3: Optimization \& Selection}
  $\mathcal{P}_{\text{final}} \gets \emptyset$ \\
  \ForEach{$p \in P_{\text{hybrid}}$}{
    $p' \gets \textsc{OptimizePrompt}(p, D_{\text{opt}})$ \tcp{Refine on Optimization Set}
    $\text{Score} \gets \textsc{Evaluate}(p', D_{\text{val}})$ \tcp{Validate on Held-Out Set}
    $\mathcal{P}_{\text{final}}.\text{append}((p', \text{Score}))$
  }
  
  $P^* \gets \arg\max_{p} (\mathcal{P}_{\text{final}})$ \\
  \Return{$P^*$}
}
\end{algorithm}

\section{Experimental Settings}

\subsection{Datasets}
To evaluate the cross-domain robustness of our framework, we utilize three real-world datasets characterized by distinct semantic structures:
\begin{itemize}
    \item MovieLens-1M (ML-1M)\footnote{\url{https://grouplens.org/datasets/movielens/}}: A dense dataset of movie ratings, representing a standard recommendation scenario with short, recognizable item titles.
    \item Amazon Games~\cite{ni-etal-2019-justifying}: A sparse dataset of video game reviews. This domain challenges the model with complex nomenclature (e.g., \textit{"Gold Edition"}, \textit{"Xbox 360"}).
    \item Bundle~\cite{10.1145/3477495.3531904}: A composite dataset spanning Electronics, Clothing, and Food. This domain introduces high ambiguity with numerical attributes (e.g., \textit{"Pack of 48"}).
\end{itemize}
Table~\ref{tab:dataset1} summarizes the statistics. For valid reproduction, we follow the strict chronological splitting protocol (8:1:1) used in the original PO4ISR~\cite{sun2024large} study.

\begin{table}[h]
\centering
\caption{Dataset statistics used for Reproducibility Benchmarking.}
\label{tab:dataset1}
\small
\begin{tabular}{l c c c}
\toprule
\textbf{Metric} & \textbf{ML-1M} & \textbf{Games} & \textbf{Bundle} \\
\midrule
\# Items & 3,416 & 17,389 & 14,240 \\
\# Sessions & 784,860 & 100,018 & 2,376 \\
Avg. Length & 6.85 & 4.18 & 6.73 \\
Sparsity & Dense & High & Medium \\
\bottomrule
\end{tabular}
\end{table}

\subsection{Implementation \& Reproducibility Parameters}
We utilize Gemini 2.0 Flash\footnote{\url{https://deepmind.google/technologies/gemini/}} as the backbone reasoning engine. To ensure fair reproducibility, we adhere to strict hyperparameter constraints. For prompt optimization, we conduct 50 training steps using a beam width of 5 and a candidate pool size of $|\mathcal{C}| = 20$. For evaluation, we report Hit Ratio (HR@k) and Normalized Discounted Cumulative Gain (NDCG@k) at $k \in {1, 5}$. To ensure robustness and eliminate stochastic variability, unlike the baseline approach, we set the LLM temperature to 0.0 and enforce deterministic output parsing.

\section{Results and Discussion}
We evaluate the reproducibility and robustness of PO4ISR$^{++}$ across three diverse datasets: ML-1M, Games, and Bundle. Table~\ref{tab:main-results} presents the comparative performance of PO4ISR$^{++}$ against a suite of baselines- MostPop, SKNN~\cite{jannach2017recurrent}, FPMC~\cite{rendle2010factorizing}, single-intent session-based methods (NARM~\cite{li2017neural}, STAMP~\cite{liu2018stamp}), multi-intent methods (GCE-GNN~\cite{wang2020global}, MCPRN~\cite{liu2023modeling}, HIDE~\cite{li2022enhancing}), and existing LLM-based reasoning methods (NIR~\cite{wang2023zero}, PO4ISR~\cite{sun2024large}, and LLM4RSR).

\subsection{The Reliability Gap vs. The Robustness Jump}
Our results reveal a distinct stability gap in the baseline LLM-based approach, which our proposed framework effectively mitigates. On ML-1M (Standard Robustness), a relatively clean dataset characterized by short, distinct movie titles, PO4ISR$^{++}$ achieves a solid improvement of 20.0\% in HR@1 ($0.2000 \rightarrow 0.2400$). This indicates that while the baseline PO4ISR is functional on standard tasks, the proposed reflexive prompting still extracts better ranking signals.

The performance gap becomes even more pronounced in the Games domain (Numeric Correction), where item titles frequently contain potentially ambiguous numerals (e.g., \textit{"Xbox 360"}, \textit{"PS4"}). Here, PO4ISR$^{++}$ shown a considerable 54.6\% improvement in HR@1 ($0.2588 \rightarrow 0.4000$). This gain is primarily driven by the Index-Based Output formatting, which prevents the LLM from misinterpreting platform version numbers as ranking indices.

The most significant improvement is observed in the Bundle dataset (Structural Correction), where PO4ISR$^{++}$ delivers a 96.4\% increase in HR@1($0.1697 \rightarrow 0.3333$). Our diagnostic analysis confirms that the baseline PO4ISR suffers near-catastrophic failure here due to ``Quantity Hallucination'' (e.g., interpreting \textit{"Pack of 48"} as \textit{"Rank 48"}). By enforcing deterministic output formats, PO4ISR$^{++}$ recovers these valid predictions that were previously discarded as parsing errors.

\begin{table*}
\centering
\caption{Reproducibility Benchmark: Comparison of PO4ISR$^{++}$ against Baselines. The massive gains in Bundle and Games highlight the correction of stochastic failures in the baseline. Best results are bolded; `*' indicates the previous SOTA.}
\label{tab:main-results}
\resizebox{\textwidth}{!}{
\begin{tabular}{l|l|ccc|cc|ccc|c|c|c|c|ccc|c}
\toprule
\textbf{Data} & \textbf{Metrics} 
& \multicolumn{3}{c|}{\textbf{Conventional}} 
& \multicolumn{2}{c|}{\textbf{Single-Intent}} 
& \multicolumn{3}{c|}{\textbf{Multi-Intent}} 
& \multicolumn{4}{c|}{\textbf{LLM Methods}} 
& \textbf{Gain} \\
\cmidrule{3-14}
& 
& MostPop & SKNN  & FPMC
& NARM  & STAMP
& GCE-GNN & MCPRN &  HIDE
&  NIR &  PO4ISR  & LLM4RSR & \textbf{PO4ISR$^{++}$}
& & \\
\midrule
\multirow{4}{*}{ML-1M} 
& HR@1      & 0.0004 & 0.1270 & 0.1132 & 0.1692 & 0.1584 & 0.1434 & 0.1498 & 0.1490 & 0.0572 & 0.2000* &   -- & \textbf{0.2400} & +20.00\% \\
& HR@5      & 0.0070 & 0.3600 & 0.3748 & 0.5230 & 0.5078 & 0.4788 & 0.4998 & 0.4932 & 0.2326 & 0.5510* & 0.2492 & \textbf{0.5900} & +7.07\%  \\
& NDCG@1    & 0.0004 & 0.1270 & 0.1132 & 0.1692 & 0.1584 & 0.1434 & 0.1498 & 0.1490 & 0.0572 & 0.2000* & -- & \textbf{0.2400} & +20.00\%\\
& NDCG@5    & 0.0053 & 0.2530 & 0.2464 & 0.3501 & 0.3367 & 0.3157 & 0.3256 & 0.3216 & 0.1436 & 0.3810* &  0.1737 & \textbf{0.4299} & +12.83\% \\
\midrule
\multirow{4}{*}{Games} 
& HR@1      & --      & 0.0020 & 0.0498 & 0.0572 & 0.0556 & 0.0522 & 0.0696 & 0.0530 & 0.1168 & 0.2588* & -- & \textbf{0.4000} & +54.56\%  \\
& HR@5      & --      & 0.0020 & 0.2564 & 0.2574 & 0.2586 & 0.2416 & 0.2694 & 0.2472 & 0.3406 & 0.5866* & 0.0596 & \textbf{0.6000} & +2.28\% \\
& NDCG@1    & --      & 0.0020 & 0.0498 & 0.0572 & 0.0556 & 0.0522 & 0.0696 & 0.0530 & 0.1168 & 0.2588* & -- & \textbf{0.4000} & +54.56\%  \\
& NDCG@5    & --      & 0.0020 & 0.1508 & 0.1534 & 0.1555 & 0.1432 & 0.1662 & 0.1475 & 0.2310 & 0.4313* & 0.0473 & \textbf{0.4994} & +15.79\% \\
\midrule
\multirow{4}{*}{Bundle} 
& HR@1      & --      & --      & 0.0398 & 0.0322 & 0.0365 & 0.0360 & 0.0458 & 0.0525 & 0.0975 & 0.1697* & -- & \textbf{0.3333} & +96.40\%  \\
& HR@5      & 0.0042 & --      & 0.2475 & 0.2322 & 0.2352 & 0.2352 & 0.2585 & 0.2644 & 0.2832 & 0.4328* & -- & \textbf{0.6000} & +38.63\%  \\
& NDCG@1    & --      & --      & 0.0398 & 0.0322 & 0.0365 & 0.0360 & 0.0458 & 0.0525 & 0.0975 & 0.1697* & -- & \textbf{0.3333} & +96.40\%  \\
& NDCG@5    & 0.0021 & --      & 0.1395 & 0.1303 & 0.1339 & 0.1267 & 0.1490 & 0.1549 & 0.1939 & 0.3040* & -- & \textbf{0.4648} & +52.89\%  \\
\bottomrule
\end{tabular}}
\end{table*}

\subsection{Ablation Study: Decomposing the Improvements}
To isolate the sources of improvement, we perform an ablation study comparing the baseline PO4ISR, our Index-Based variants (Prompt 1/2), and the full Domain-Specific optimization (Table~\ref{tab:ablation}). The results highlight two critical insights for reproducibility. First, the impact of index-based stability is clearly evident. In the Games dataset, simply switching to an Index-Based output (Prompt 1) without changing the reasoning logic boosts HR@1 from $0.260 \rightarrow 0.350$. This confirms that the ``reasoning'' was correct, but the ``parsing'' was the bottleneck. Second, the contrast between domain specificity and cross-domain fusion further emphasizes the importance of adaptive strategies. While Prompt 2 excels in dense domains (ML-1M: HR@1 0.260), it falters in sparse ones. Conversely, Prompt 1 dominates Games. This variance underscores the need for Reflexive Cross-Domain Fusion, which dynamically selects the optimal reasoning path rather than relying on a static ``Top-1'' prompt.

\begin{table}[h]
\centering
\caption{Ablation analysis. Index-Based formatting alone provides a significant boost in sparse domains (Games), while domain-specific optimization is crucial for complex domains (Bundle).}
\label{tab:ablation}
\resizebox{\linewidth}{!}{
\begin{tabular}{l|cc|cc|cc}
\toprule
\textbf{Method} & \multicolumn{2}{c|}{\textbf{ML-1M}} & \multicolumn{2}{c|}{\textbf{Games}} & \multicolumn{2}{c}{\textbf{Bundle}} \\
& HR@1 & NDCG@5 & HR@1 & NDCG@5 & HR@1 & NDCG@5 \\
\midrule
Initial Baseline & 0.143 & 0.2823 & 0.079 & 0.2110 & 0.050 & 0.1396 \\
PO4ISR (Original) & 0.211 & 0.3975 & 0.260 & 0.4381 & \textbf{0.193} & 0.3183 \\
\midrule
\textbf{+ Index-Based (P1)} & 0.150 & 0.3329 & \textbf{0.350} & \textbf{0.4713} & 0.170 & 0.3273 \\
\textbf{+ Index-Based (P2)} & \textbf{0.260} & \textbf{0.4336} & 0.333 & 0.4042 & 0.140 & 0.2852 \\
\textbf{+ Domain Specific} & 0.230 & 0.3799 & 0.260 & 0.3826 & 0.190 & \textbf{0.3553} \\
\bottomrule
\end{tabular}
}
\end{table}

\subsection{Diagnosing the Failure Modes}
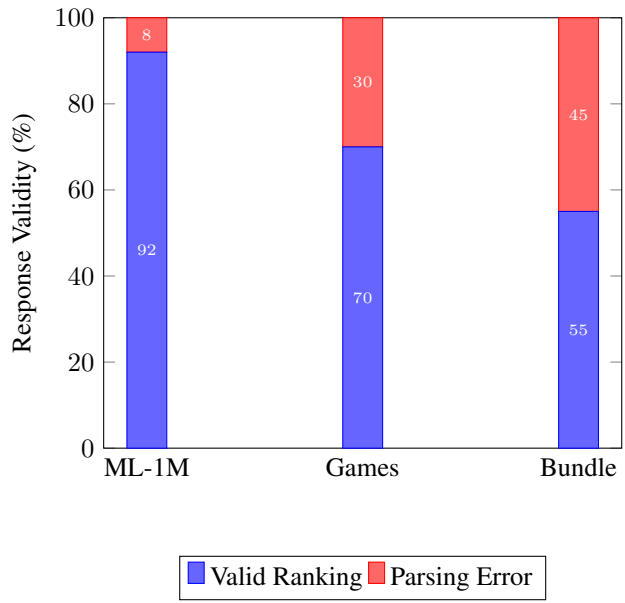
\begin{figure}[h]
\centering
\begin{tikzpicture}
\begin{axis}[
    ybar stacked,
    bar width=15pt,
    symbolic x coords={ML-1M, Games, Bundle},
    xtick=data,
    ylabel={Response Validity (\%)},
    legend style={at={(0.5,-0.25)},anchor=north,legend columns=-1},
    ymin=0, ymax=100,
    nodes near coords,
    every node near coord/.append style={font=\tiny, color=white}
]
    \addplot+[fill=blue!60] coordinates {(ML-1M,92) (Games,70) (Bundle,55)};
    \addplot+[fill=red!60] coordinates {(ML-1M,8) (Games,30) (Bundle,45)};
    \legend{Valid Ranking, Parsing Error}
\end{axis}
\end{tikzpicture}
\caption{\textbf{Diagnosing the stability gap:} The distribution of valid vs. unparseable outputs in the baseline PO4ISR. The significant performance drop in Games/Bundle correlates directly with a high rate of parsing errors (Red), which PO4ISR$^{++}$ eliminates via deterministic indexing.}
\label{fig:parsing_error}
\end{figure}

To further validate our robustness claims, we visualize the stability gap in Figure~\ref{fig:parsing_error}. The stacked bar chart illustrates the proportion of Invalid/Unparseable responses in the baseline model. Qualitative error analysis of 100 baseline failures shows three distinct ``hallucination'' categories that PO4ISR$^{++}$ resolves:
\begin{itemize}
    \item Numerical Confusion (45\%): In Bundle, items like \textit{"Goya Mango Nectar (Pack of 48)"} were systematically misranked as "Rank 48". 
    \item Platform Ambiguity (30\%): In Games, distinctions between \textit{"Xbox 360"} and \textit{"Xbox One"} were often lost in text generation. The reflexive prompt fusion retains these nuances by leveraging domain-specific descriptors.
    \item Short-Title Noise (25\%): In ML-1M, generic titles like \textit{"Speed"} caused context loss, which the reflexive prompting mitigates by enforcing structured reasoning.
\end{itemize}

\subsection{Cross-Domain Generalization}
\begin{figure}[h]
\centering
\begin{tikzpicture}
\begin{polaraxis}[
    ymax=0.65,
    xtick={0,120,240},
    xticklabels={ML-1M (HR@5), Games (HR@5), Bundle (HR@5)},
    grid=both,
    major grid style={dotted, gray},
    legend style={at={(0.5,-0.2)},anchor=north},
]
    \addplot[mark=*, thick, color=red, fill=red!20, fill opacity=0.5] 
    coordinates {(90,0.5510) (210,0.5866) (330,0.4328) (90,0.5510)};
    \addlegendentry{PO4ISR (Baseline)}
    
    \addplot[mark=square*, thick, color=blue, fill=blue!20, fill opacity=0.5] 
    coordinates {(90,0.5900) (210,0.6000) (330,0.6000) (90,0.5900)};
    \addlegendentry{PO4ISR$^{++}$}
\end{polaraxis}
\end{tikzpicture}
\caption{\textbf{Robustness Profile:} While the baseline (Red) struggles to generalize to the complex Bundle domain, PO4ISR$^{++}$ (Blue) maintains a consistent high-performance envelope across all three semantic domains.}
\label{fig:radar_robustness}
\end{figure}
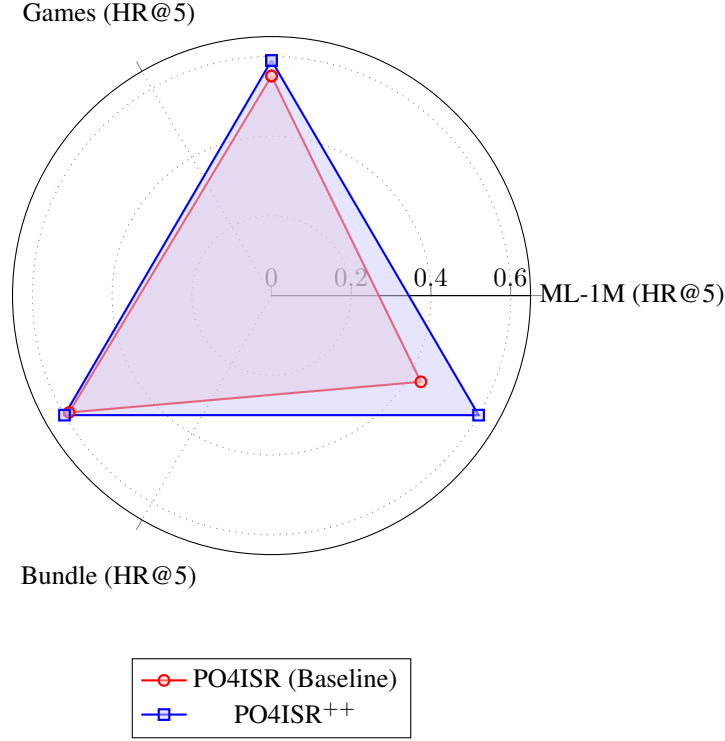

Finally, we demonstrate the robustness of the reflexive fusion strategy in Figure~\ref{fig:radar_robustness}. Unlike the baseline, which overfits to the standard ML-1M domain, PO4ISR$^{++}$ maintains a consistent high-performance envelope across all three semantic contexts.

The superior performance of PO4ISR$^{++}$ is not merely a result of better training, but of structural decoupling. By separating the ranking index from the semantic content and by employing the reflexive fusion to adapt to domain-specific logic, we provide a reproducible framework that is resilient to the stochastic nature of LLMs.

\section{Conclusion}
In this work, we conducted a rigorous reproducibility study of the reasoning-based PO4ISR framework. Our analysis uncovered a stability gap where standard text-based prompting fails in domains with numerical ambiguity (Games, Bundle). We introduced PO4ISR$^{++}$, a robust extension that enforces Deterministic Index Mapping and employs Reflexive Cross-Domain Fusion. The experiments demonstrate that these interventions do not merely improve performance but also fundamentally restore the recommender's reliability. We achieve stabilized gains of up to 96\% in HR@1 on complex datasets, proving that LLM-based recommendation can be both high-performing and reproducible. We will release our audit logs and code to facilitate future robustness research.


\bibliographystyle{unsrtnat}
\bibliography{sample-base}

\appendix

\section{Prompts}
\label{sec:prompt}
The prompts used for the experiments are listed here.

{\scriptsize
\begin{tcolorbox}[
    colback=blue!5!white,
    colframe=blue!75!white,
    title=\textbf{\textcolor{white}{Prompt 1: Task Description}},
    fonttitle=\bfseries\itshape,
    coltitle=black,
    boxrule=0.8pt,
    arc=3pt,
    left=5pt,
    right=5pt,
    top=5pt,
    bottom=5pt
]
\label{init} \textit{Based on the user's current session interactions, you need to answer the following subtasks step by step:\\
Identify combinations of items within the session, where the number of combinations can be one or more.\\
Infer the user's intent for each combination based on the items involved.\\
Select the most relevant intent that best represents the user's current preferences.\\
Rerank the items in the candidate set (provided below) based on the selected intent, considering the likelihood of potential user interactions.\\
Return the indices of the reranked items in the same order as they appear in the reranked list, ensuring that the indices (0 based indexing) refer to their original position in the candidate set. \\
Return Only the indices enclosed in for example [3,10,1......] nothing else. No explanations no nothing.}
\end{tcolorbox}

\begin{tcolorbox}[
    colback=blue!5!white,
    colframe=blue!75!white,
    title=\textbf{\textcolor{white}{Prompt 2: Input Data \citet{sun2024large}}},
    fonttitle=\bfseries\itshape,
    coltitle=black,
    boxrule=0.8pt,
    arc=3pt,
    left=5pt,
    right=5pt,
    top=5pt,
    bottom=5pt
]
\textit{Current session interactions: \{\texttt{[idx:"item title", ...]}\} \\
Candidate item set: \{\texttt{[idx:"item title", ...]}}
\end{tcolorbox}

\begin{tcolorbox}[
    colback=blue!5!white,
    colframe=blue!75!white,
    title=\textbf{\textcolor{white}{Prompt 3: Inferring Reasons for Errors~\cite{sun2024large}}},
    fonttitle=\bfseries\itshape,
    coltitle=black,
    boxrule=0.8pt,
    arc=3pt,
    left=5pt,
    right=5pt,
    top=5pt,
    bottom=5pt
]
\label{reason} \textit{I'm trying to write a zero-shot recommender prompt. \\
My current prompt is \{\texttt{prompt}\}. \\
But this prompt gets the following example wrong: \{\texttt{error\_case}\}, give \{\texttt{N\_r}\} reasons why the prompt could have gotten this example wrong. \\
Wrap each reason with \texttt{<START>} and \texttt{<END>}.}
\end{tcolorbox}



\begin{tcolorbox}[
    colback=blue!5!white,
    colframe=blue!75!white,
    title=\textbf{\textcolor{white}{Prompt 4: Refining Prompts with Reasons~\cite{sun2024large}}},
    fonttitle=\bfseries\itshape,
    coltitle=black,
    boxrule=0.8pt,
    arc=3pt,
    left=5pt,
    right=5pt,
    top=5pt,
    bottom=5pt
]
\label{opt} \textit{I'm trying to write a zero-shot recommender prompt.\\
My current prompt is \{\texttt{prompt}\}.\\
But this prompt gets the following example wrong: \{\texttt{error\_case}\}. \\
Based on the example, the problem with this prompt is that \{\texttt{reasons}\}. \\
\label{opt} Based on the above information, please write one improved prompt. At the end, necessarily include this as it is - 'Your answer should only be ``[8,1,6,...]'', give me only indices in list format, nothing else, no explanations no nothing, only and only list of indices and remember to rank all 20 candidates in the candidate\_set (I want all indices from 0 to 19). The prompt is wrapped with \texttt{<START>} and \texttt{<END>}. \\
The new prompt is:}
\end{tcolorbox}

\begin{tcolorbox}[
    colback=blue!5!white,
    colframe=blue!75!white,
    title=\textbf{\textcolor{white}{Prompt 6: Augmenting Prompts~\cite{sun2024large}}},
    fonttitle=\bfseries\itshape,
    coltitle=black,
    boxrule=0.8pt,
    arc=3pt,
    left=5pt,
    right=5pt,
    top=5pt,
    bottom=5pt
]
\label{aug} \textit{Generate a variation of the following instruction while keeping the semantic meaning.\\
At the end, necessarily include this as it is - 'Your answer should only be ``[8,1,6,......]'', give me only indices in list format, nothing else, no explanations no nothing, only and only list of indices and remember to rank all 20 candidates in the candidate\_set (I want all indices from 0 to 19).\\
``Input: $refined\_prompt$''\\
``Output:''
}
\end{tcolorbox}

\begin{tcolorbox}[
    colback=blue!5!white,
    colframe=blue!75!white,
    title=\textbf{\textcolor{white}{Prompt-1-ml - (one of the best prompt from ml according~\cite{sun2024large})}},
    fonttitle=\bfseries\itshape,
    coltitle=black,
    boxrule=0.8pt,
    arc=3pt,
    left=5pt,
    right=5pt,
    top=5pt,
    bottom=5pt
]
\textit{Given the user's current session interactions, please follow these steps to complete the subtasks:\\
1. Identify potential patterns or themes within the session by analyzing combinations of items.\\
2. Consider additional contextual information, such as genre, director, or actor, to better understand the user's interactive intent.\\
Evaluate the inferred intents alongside the user's previous ratings and preferences to determine the most accurate representation of their current preferences.\\
Utilize the selected intent to rank the 20 items in the candidate set based on the likelihood of user interactions. Provide the ranking results with the corresponding item index.\\
Ensure that the order of all items in the candidate set is provided, and that the ranking items are within the candidate set.}
\end{tcolorbox}

\begin{tcolorbox}[
    colback=blue!5!white,
    colframe=blue!75!white,
    title=\textbf{\textcolor{white}{Prompt-1-games - (one of the best prompt from games according~ \cite{sun2024large})}},
    fonttitle=\bfseries\itshape,
    coltitle=black,
    boxrule=0.8pt,
    arc=3pt,
    left=5pt,
    right=5pt,
    top=5pt,
    bottom=5pt
]
\textit{Your task is to analyze the user's current session interactions and the candidate set of items to accurately infer the user's preferences and intent.\\
1. Consider any patterns or combinations of items within the session that may indicate the user's genre preference or other relevant criteria.\\
2. Evaluate the context and relevance of the items in the candidate set to the user's session interactions, taking into account factors like genre, price, or other relevant criteria.\\
3. Deduce the user's interactive intent within each combination, considering factors like price comparison, genre preference, or other relevant criteria.\\
4. Rearrange the items in the candidate set based on the inferred intent, ensuring that the items are ordered according to the likelihood of potential user interactions.\\
Provide the rearranged list of items from the candidate set, along with their corresponding item indices.}
\end{tcolorbox}

\begin{tcolorbox}[
    colback=blue!5!white,
    colframe=blue!75!white,
    title=\textbf{\textcolor{white}{Opt prompt bundle - (Prompt optimized by the proposed approach)}},
    fonttitle=\bfseries\itshape,
    coltitle=black,
    boxrule=0.8pt,
    arc=3pt,
    left=5pt,
    right=5pt,
    top=5pt,
    bottom=5pt
]
\textit{1. \textbf{Session Understanding and Intent Disambiguation:} Analyze the user\'s session interactions to identify the most prominent themes or categories. Prioritize themes represented by multiple items or those with more specific product descriptions. In cases of ambiguity, lean towards broader categories (e.g., ``Electronics Accessories'' instead of overly specific ones like ``Surveillance Equipment''). The aim is to capture the overarching need driving the user\'s interaction.\\
2. \textbf{Category Representation:} Create a vector representation for both the inferred session category (from Step 1) and each item in the candidate set. Use product descriptions and, if available, category metadata to populate these vectors. Consider using techniques like TF-IDF or embedding models (though you won\'t execute these directly, the concept guides the scoring).\\
3. \textbf{Relevance Scoring (Cosine Similarity):} Calculate the cosine similarity between the session category vector and each item\'s vector. This provides a numerical relevance score representing the semantic similarity between the user\'s session and each candidate item.\\
4. \textbf{Attribute Weighting (Enhancement, not Replacement):} Identify key attributes from the session interactions (e.g., ``USB,'' ``battery powered,'' ``case''). Increase the relevance score of items that explicitly possess these attributes, but only as a multiplier to the base cosine similarity score (e.g., 1.1x for each matching attribute). Avoid relying solely on attribute matching; the cosine similarity should remain the primary driver. \\
5. \textbf{Negative Scoring (Strong Penalties):} Implement strong negative scores for items that are demonstrably dissimilar to the overall session. For instance, if the session revolves around electronics, assign a high negative score to clothing or food items. Establish a clear threshold for ``dissimilarity'' based on keyword analysis (e.g., fewer than 2 shared keywords with any item in the session).\\
6. \textbf{Diversity Consideration (Subtle Adjustment):} After the initial ranking, slightly penalize items that are extremely similar to each other within the top ranks. This encourages some diversity in the recommendations. This is a minor adjustment (e.g., -0.01 to the score of a very similar item). Only apply this to top 5 ranked items and only if cosine similarity between them is >0.9.\\
7. \textbf{Ranking:} Rank the items in the candidate set based on their final adjusted relevance scores ($Cosine\;Similarity + Attribute\;Weighting - Negative\;Scoring - Diversity\;Adjustment$), from highest to lowest.\\
8. \textbf{Index Mapping:} Provide the ranking results by listing the item indices (starting from 0) corresponding to the reordered candidate set. Ensure all 20 items from the candidate set are included in the ranked list.\\ Your answer should only be [8,1,6,...], give me only indices in list format nothing else, no explanations no nothing only and only list of indices and remember, rank all 20 candidates in the candidate set (I want all indices from 0 to 19).}
\end{tcolorbox}

\begin{tcolorbox}[
    colback=blue!5!white,
    colframe=blue!75!white,
    title=\textbf{\textcolor{white}{Prompt 8: Combining Prompts}},
    fonttitle=\bfseries\itshape,
    coltitle=black,
    boxrule=0.8pt,
    arc=3pt,
    left=5pt,
    right=5pt,
    top=5pt,
    bottom=5pt
]
\textit{Generate 10 different prompts, combining the strengths of 3 different prompts, so that these 10 can perform on all 3 domains. Give me no other text, only a list of prompts like this [prompt1, prompt2, $\ldots$, prompt10]. I want no other text only this list. Each prompt should be one line and can have $\geq500$ number of words, and write this as last line of each generated prompt -\textit{`Give me only ranked indices of candidate set in list format nothing else, no explanations no nothing only and only list of indices and remember, rank all $20$ candidates in the candidate set (I want all indices from $0$ to $19$), for example if $8^{th}$ element in candidate set (from left) should be top ranked, then 1st then $14^{th}\ldots$ return their $0$ based indexes like $[7,0,13,\ldots]$ and rank all $20$'}}
\end{tcolorbox}

\begin{tcolorbox}[
    colback=red!5!white,
    colframe=red!75!white,
    title=\textbf{\textcolor{white}{Best Prompt}},
    fonttitle=\bfseries\itshape,
    coltitle=black,
    boxrule=0.8pt,
    arc=3pt,
    left=5pt,
    right=5pt,
    top=5pt,
    bottom=5pt
]
\textit{Prioritize positive associations and balance negative penalties when ranking candidate items against a user's session. Analyze the user's session to identify key themes, considering interaction frequency as a signal of importance, with more recent interactions carrying greater weight. Infer user intent by identifying dominant product categories (e.g., snacks, meal ingredients, beverages) and sub-themes, utilizing TF-IDF, word embeddings, or large language models to represent both session interactions and candidate items as vectors capturing semantic meaning. When uncertain about narrow themes, broaden the scope to more general categories, such as ``food,'' ``pantry staples,'' or ``beverages.'' Calculate cosine similarity between the session theme vector and each item vector to quantify semantic relevance. Boost the relevance score of items with explicit attribute matches (e.g., ``apple pie'' if ``LARABAR Apple Pie'' is in the session), but avoid penalizing candidate items that are related, like apple sauce. Furthermore, add a multiplier based on how similar the candidate item\'s broader category is to the session\'s most frequent broader category. Negatively score candidate items only if they are categorically dissimilar AND contain keywords directly contradictory to the user's session. Implement a diversification penalty by slightly reducing the relevance score of highly similar top-ranked items. Explicitly promote direct matches present in both the session and candidate set. Consider co-occurrence patterns in the session, for example, items frequently purchased together, and apply positive biases to similar combinations in the candidate set. Factor in interaction recency, with more recent interactions having a stronger influence on inferred user intent. Ensure that the order of all items in the candidate set is provided, and that the items for ranking are within the candidate set. Order the candidate items by their adjusted relevance scores, reflecting the likelihood of user interaction. Evaluate the context and relevance of the items in the candidate set to the user\'s session interactions, taking into account factors like genre, price, or other relevant criteria; Evaluate candidate items for how well they might complement the items in the user\'s session and boost items that represent complements. Your answer should only be [8,1,6,...], give me only indices in list format nothing else, no explanations no nothing only and only list of indices and remember, rank all 20 candidates in the candidate set (I want all indices from 0 to 19).}
\end{tcolorbox}
}

\end{document}